\documentclass[12pt,preprint]{aastex}

\slugcomment{}

\shorttitle{Subaru Image for HR~8799 in 2002}
\shortauthors{Fukagawa et al.}

\def\II{I~\hskip -\lastskip \hspace{-.1em}I\ }

\begin{document}


\title{$H$ band Image of a Planetary Companion around HR~8799 in 2002 \footnotemark[1]}


\author{Misato Fukagawa\altaffilmark{2}, Yoichi Itoh\altaffilmark{3}, Motohide Tamura\altaffilmark{4, 5}, Yumiko Oasa\altaffilmark{3}, Saeko S. Hayashi\altaffilmark{5, 6}, Yutaka Fujita\altaffilmark{2}, Hiroshi Shibai\altaffilmark{2} and Masahiko Hayashi\altaffilmark{5, 6}} 

\altaffiltext{2}{Department of Earth and Space Science, Graduate School of Science, Osaka University, 1-1 Machikaneyama, Toyonaka, Osaka 560-0043 Japan}  
\altaffiltext{3}{Graduate School of Science, Kobe University, 1-1 Rokkodai, Nada, Kobe, Hyogo 657-8501, Japan} 
\altaffiltext{4}{Department of Astronomical Science, The Graduate University for Advanced Studies (SOKENDAI), 2-21-1 Osawa, Mitaka, Tokyo 181-8588, Japan}   
\altaffiltext{5}{National Astronomical Observatory of Japan, 2-21-1 Osawa, Mitaka, Tokyo 181-8588, Japan} 
\altaffiltext{6}{Subaru Telescope, National Astronomical Observatory of Japan, 650 North A'ohoku Place, Hilo, HI 96720, USA}

\footnotetext[1]{Based on data collected at the Subaru Telescope, which
is operated by the National Astronomical Observatory of Japan.}

\begin{abstract}
The discovery of three planetary companions around HR~8799 \citep{mar08} marked a significant epoch in direct imaging of extrasolar planets. Given the importance of this system, we re-analyzed $H$ band images of HR~8799 obtained with the Subaru 36-elements adaptive optics (AO) in July 2002. The low-order AO imaging combined with the classical PSF-subtraction methods even revealed the extrasolar planet, HR~8799b. Our observations in 2002 confirmed that it has been orbiting HR~8799 in a counter-clockwise direction. The flux of HR~8799b was consistent with those in the later epochs within the uncertainty of 0.25~mag, further supporting the planetary mass estimate by \cite{mar08}. 
\end{abstract}



\section{Introduction}
The number of extrasolar planets discovered is increasing but primarily with indirect methods such as measurements of radial velocity attributed to the wobble of parent stars \citep{mar05,may04}. The detections of photons from extrasolar planets have been achieved with {\it Spitzer Space Telescope} that captured the thermal infrared light during their secondary transits \citep{cha05}. Direct imaging, in which the light from planets can be spatially separated from the host star, has long been desired since it opens up detailed characterization of the planets. The planetary-mass companion candidates have indeed been imaged around several young pre-main- or main-sequence stars \citep{lag09,kal08}, but the derivation of the planetary mass is often model-dependent or it is still uncertain if the companion candidate is gravitationally bound to the host star  \citep[e.g.,][]{cha04,neu05,laf08}. 

Most recently, \cite{mar08} reported one of the robust and exciting results: the discovery of a planetary system around an A5 star HR~8799  with direct imaging. The three Jupiter-like planets, HR~8799b, c, and d were detected at the separations relative to the star of $1\farcs72$, $0\farcs96$, and $0\farcs62$, respectively. Their multi-epoch astrometry from 2004 to 2008 clearly shows that the companion candidates are not background objects but rather orbiting the central star in a counter-clockwise direction, probably moving in a nearly circular orbit. It is worth noting that their results in 2007 and 2008, that led to the significant detections of HR~8799c and d, were obtained using the angular differential imaging technique that effectively suppressed the speckle noise. The host star, HR~8799 (HD~218396), has been recognized as a Vega-like star \citep{syl96}.  Although the infrared excess is not quite large ($L_{\rm{IR}}/L_{*} = 2.0 \times 10^{-4}$), its proximity ($d = 39.4$~pc, \citealt{lee07}) is suited to resolve the structure in the inner part, and hence for direct imaging. Therefore, we coronagraphically observed it in 2002 with the Subaru Telescope, originally intended to find scattered light from the debris disk. Following the discovery by \cite{mar08}, we revisited our Subaru data and applied thorough analysis suited for point source detection, 
 since the astrometry in other epochs would have an importance for future accurate determinations of the planetary orbits, as well as for independent confirmation of the planetary companions. In the present $Letter$, we report the detection of HR~8799b in 2002. 

\section{Observations}
HR~8799 ($H=5.28$) was observed on 2002 July 19 (UT) with the near-infrared coronagraphic camera CIAO (Coronagraphic Imager with Adaptive Optics; \citealt{tam00}) mounted on the Subaru Telescope. The images were obtained in the $H$ band ($\lambda=1.65~\micron$) with the Subaru  36-elements adaptive optics \citep{tak04}. The $0\farcs5$ diameter transmissive coronagraphic mask and the 80\% diameter Lyot stop without spider vanes were used. 
 The sky was clear and the seeing was relatively good ($\sim0\farcs6$ at optical) and stable except for at the beginning of the observations. No special technique optimized for planet detection was applied such as angular or spectral differential imaging. CIAO is equipped with a 1024$\times$1024 InSb ALADDIN \II array. Since our observations were aimed to detect faint extended emission from the debris disk, the medium-resolution mode (21~mas pixel$^{-1}$) was used to obtain the better sensitivity. Nevertheless, the pixel scale of 21~mas was sufficiently small to measure the positional difference for the companion candidates around HR~8799 given their large orbital motions (25~mas/yr for the most distant companion; \citealt{mar08}). The pixel scale and orientation of the detector that we adopt here are $21\farcs2 \pm 0\farcs4$~mas~pixel~$^{-1}$ and $5\arcdeg9 \pm 0\fdg5$, respectively, derived from the observations in 2001 and 2002 of the Trapezium cluster \citep{ito05}, M15, and objects that have other nearby point sources. The distortion is negligible and was not corrected. As a PSF reference star, HD~218172 ($H=5.93$, separated by $1\arcdeg$ from HD~218396 on the projected sky) was also imaged with the same occulting mask of $0\farcs5$ just before and after observing HR~8799. The AO guide stars were HR~8799 and the reference star themselves, but they were too bright for the AO and their fluxes did not match in $R$ band where the wavefront sensing was performed. Therefore, the neutral density filters were inserted in front of the lenslet array in the AO unit, and by selecting the appropriate number of the filters for each star, the similar performance in the atmospheric correction was achieved for both sources. The 5 exposures of 1~sec were coadded into one data frame, and 264 and 200 frames were obtained in total for HR~8799 and the PSF reference star, respectively. 
The position angle of the spider pattern was varied with time in the images, since the instrument rotator was used with the altitude-azimuth telescope. HR 8799 and the PSF reference star were not observed at the meridian (near zenith), and the spider pattern was rotated only by 4 degrees during the observation of HR~8799. The stars, which were tracked at optical, were gradually misaligned with the mask center with time, and the positional adjustment was made with the AO tip-tilt mirror. 
 The focal plane mask was moved by 1$\arcsec$ and the star was placed on the mask using the AO tip-tilt after taking 132 frames for HR~8799 (dithering). 
FS~155 was observed on the same night as a flux calibrator, without the occulting mask and without the AO loop closed to minimize the observational overhead. 


\section{Data Reductions}
The obtained images were calibrated using IRAF for dark subtraction, flat-fielding with dome-flats, bad pixel substitution, and sky subtraction. The dark exposures were obtained before the observations were started, and the dome flats were taken using an incandescent lamp at the end of the night. The constant sky level was estimated for each frame measuring the median value at large separations ($>5\arcsec$ from the star). 

In order to extract the extended emission or companion candidates in the vicinity of the bright star, the PSF-subtraction was carried out. First, the careful selection of the usable data was required to reduce the effect of the  temporal variations in the PSFs. As a measure of the PSF quality, we used the brightness ratio of the halo at $r=1\farcs5$ and the peak intensity, then eliminated the images with the ratio larger than 0.02 for both HR~8799 and the PSF reference. 
The 220 frames were selected for HR~8799, corresponding to the total integration time of 18.3~minutes, and the 141 frames (11.8~minutes) were adopted for the reference star. Second, the stellar position was estimated by measuring the centroid of the transmitted stellar light in the mask, and the stars were registered to a common position. 
Each four consecutive frames were averaged to improve the signal-to-noise ratio (SNR). 

We applied the following procedures in order to remove the stellar halo. 

\begin{figure}
\plotone{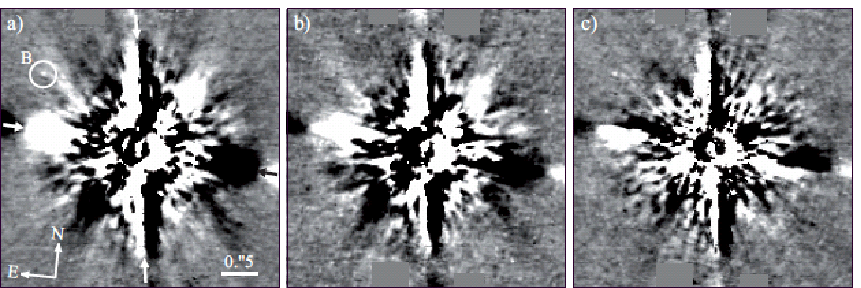}
\caption{HR~8799 in $H$ band with the rotated image subtractions. (a) The median-combined image after subtracting the self-rotated images. The strong features running in north-south and east-west directions are caused due to the spider of the secondary mirror, and denoted with arrows. (b)  The PSF reference star image with the subtraction of the rotated images. (c) The difference between the self-rotated subtractions for HR~8799 and the reference star. The field of view is 4$\arcsec \times 4\arcsec$. All the images are shown in the same display range. \label{fig:rotsub}} 
\end{figure}

\begin{figure}
\plotone{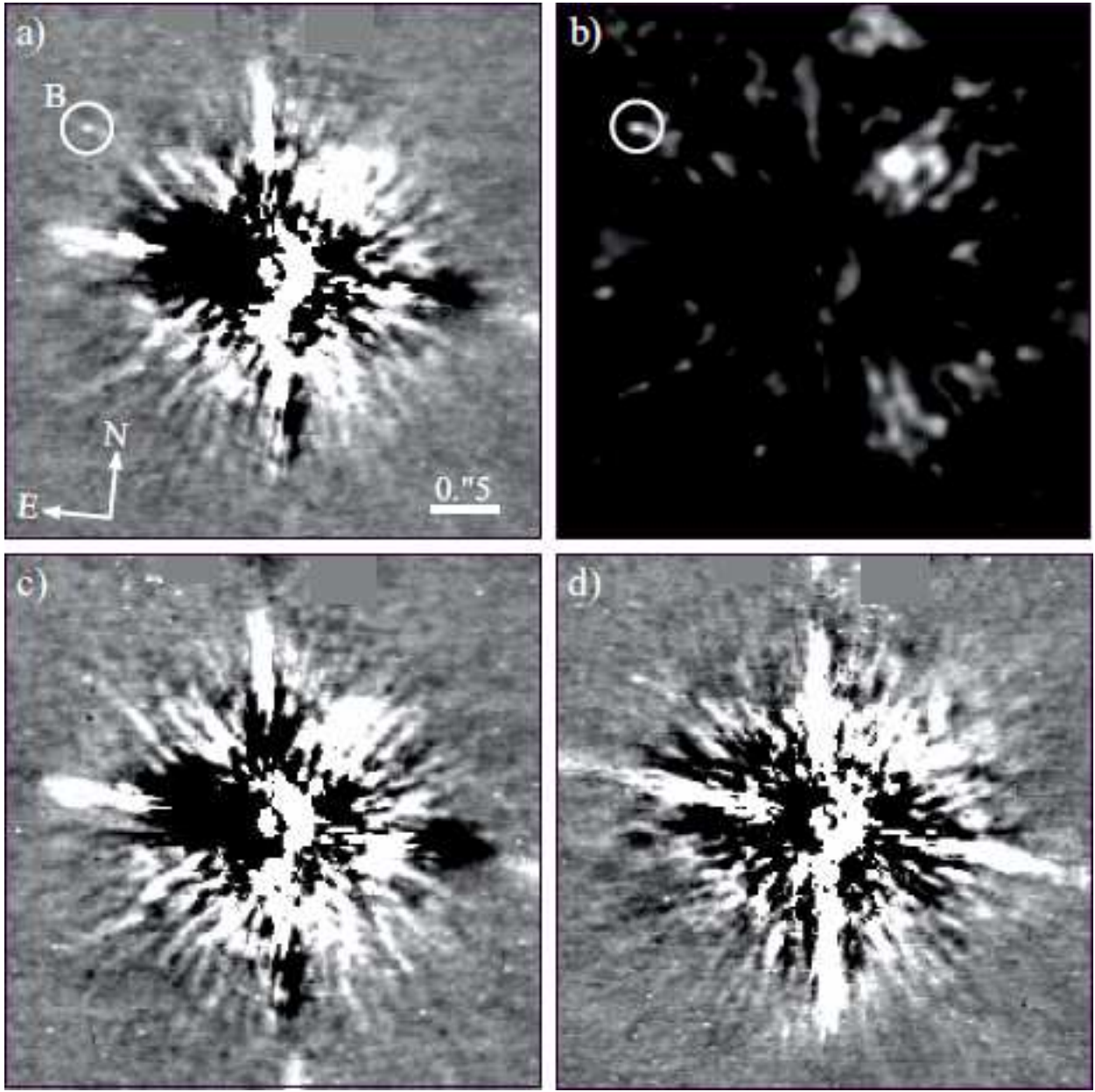}
\caption{HR~8799 in $H$ band with the subtractions of the PSF constructed from the reference star images. (a) The median-combined PSF-subtracted image.  (b) The PSF-subtracted image divided by the standard deviation of pixel values. The map was smoothed by the Gaussian function with its sigma of 1.6~pixel. (c) The median-combined images obtained before the dithering, and  (d) after the dithering. The field of view is 4$\arcsec \times 4\arcsec$. The panel (a), (c), and (d) are displayed in the same range. The panel (b) was linearly shown in the range of 0.5--1.5, but if the Poisson noise component is subtracted, the SNR for the emission B is 4. \label{fig:psfsub}} 
\end{figure}

\begin{figure}
\plotone{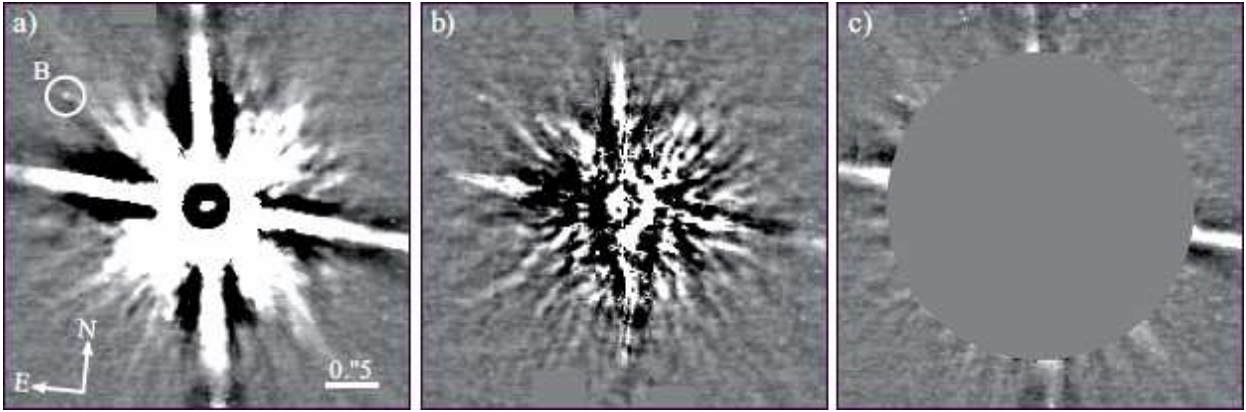}
\caption{(a) The subtractions of median-filtered self images. (b) The subtractions of the median-filtered reference images. (c) The result of the LOCI subtraction.  The field of view is 4$\arcsec \times 4\arcsec$. All the images are shown in the same display range. \label{fig:med_locisub}} 
\end{figure}

\begin{enumerate}
\item Subtraction of the image rotated by 180$\arcdeg$.\\
The rotated image for HR~8799 was subtracted from the original unrotated one, but there remained an asymmetric positive and negative pattern in every frame since the PSF was not completely circular. In order not to cause the asymmetry, we had to slightly shift the rotated image. The amount of the shifts depended on each frame, but the maximum offsets are $\Delta \alpha = 0\farcs065$ and $\Delta \delta = 0\farcs028$. The relatively large shifts do not indicate that the centroid measurement of the stellar position was incorrect, since the spider pattern was clearly misaligned in the subtracted result with the re-registered frame. However, we conservatively included the averaged offsets, $\Delta \alpha = 0\farcs021$ and $\Delta \delta = 0\farcs0005$, in the error of the astrometry for HR~8799b described in the later section. The subtraction was made for each frame of HR~8799, and the resultant frames were median combined (the  panel (a) of Figure~\ref{fig:rotsub}). Furthermore, the same subtractions using the rotated images were performed for the PSF reference star. The result  was scaled to the brightness of HR~8799, rotated to cancel the difference of the instrumental rotator angle, then subtracted from the result for HR~8799 to suppress the speckle noise (the panel (c) of Figure~\ref{fig:rotsub}). 
 
\item Subtraction of the PSF constructed from the reference star images. \\
Five PSFs were built by averaging all the reference images, images obtained before or after observing HR~8799, images that have the brightness ratio of the halo ($r=1\farcs5$) and the peak of $<0.01$, and those with the ratio of $<0.008$. The PSF subtraction was performed for each frame of HR~8799 using the averaged reference PSF with the higher SNR. The PSF was rotated to match the position angle of the spider pattern for HR~8799, and scaled to fit the halo brightness of HR~8799 at $r=1\farcs5$. It was necessary to manually tweak the scaling factor and the position of the PSF to improve the subtraction. The best PSF was chosen so that it gave the smallest residuals. The PSF-subtracted images were median combined and shown in the panel (a) of Figure~\ref{fig:psfsub}. 
The real point source should be present in all the data frames. In order to extract such emission, the standard deviation of the pixel values was calculated at each pixel position (a noise map). 
The PSF-subtracted image was divided by the noise map to emphasize the components that were temporally stable (the panel (b) of  Figure~\ref{fig:psfsub}). 
 
\item Subtraction of the median-filtered images.\\
This procedure was applied in order to remove the emission with lower spatial frequency than point sources. The accurate measurement of the FWHM for a point source was impossible since there was no bright, unocculted star in the small field of view of CIAO. However, the light from the PSF core for HR~8799 was transmitted through the occulting mask, and the derived FWHM ($0\farcs07\pm0\farcs01$) was a reasonable value for the AO36 imaging using a bright guide star under good seeing condition. The consistency of the FWHMs between unmasked companions and a masked star was also confirmed for HD~200775 observed on the same night, but under worse seeing condition, $i.e.$, the FWHM of $0\farcs1$ and the brightness ratio of the halo and the peak of $>$0.02.  
In order to extract the emission that has the similar spatial scale as a point source ($0\farcs07$ = 3.3~pixels), the median-filtered image was constructed using a median window of 21$\times$21 pixels and subtracted. The resultant images were median combined to reduce the peaky noise pixels (the panel (a) of Figure~\ref{fig:med_locisub}). The reference frames used for the reduction 2 (the subtraction of the PSF of the reference star) were also processed in the same way, and subtracted from the filtered image for HR~8799. The each subtracted images were finally median combined and displayed in Figure~\ref{fig:med_locisub}. 

\item Subtraction using the Locally Optimized Combination of Images (LOCI) algorithm. \\
\cite{laf07} proposed the effective PSF subtraction algorithm in which reference images are linearly combined with coefficients of the combination  estimated to minimize the residuals in each subsection of the image. Following \cite{laf07}, the squared residuals of the subtraction of the reference PSF ($O^R$) from the target ($O^T$) are expressed as $\sigma^2 = \sum_i (O^T_i - O^R_i)^2 = \sum_i (O^T_i - \sum_k c^k O^k_i)^2 $, where $i$ denotes a pixel in the subsection and $k$ is the index of the reference image. The coefficients for the $k$th image, $c^k$, can be calculated from linear equations, ${\bold \it A} c = {\bold \it b}$ where $A_{jk} = \sum_i O^j_i O^k_i,\ b_j = \sum_iO^j_i O^T_i$. 
The PSFs for the reduction 2 (rotated and scaled ones) were used as reference images.  Before applying the LOCI, we reduced the low-frequency emission that produced the huge mismatch of the PSFs, subtracting the median filtered image using a 21$\times$21 box to calculate the median value. 
The subtraction was made for each HR~8799 image in the region of interest ($r > 1\farcs5$), with the radial and azimuthal width of the subsection set to be $0\farcs5$ and 20$\arcdeg$, respectively (the area corresponds to 79 PSF cores with $0\farcs07$ FWHM). The median combined result is shown in the panel (c) of Figure~\ref{fig:psfsub}. We tried to apply the subtractions using smaller subsections for the inner region, but did not produce a significant detection of HR~8799c. 
\end{enumerate}

\section{Results}
The observations were originally designed to detect the light scattered from the disk. However, no extended emission was found beyond the boundary where the subtraction residual started to dominate ($r \lesssim1\farcs4$). The detection limit is 16~mag~arcsec$^{-2}$ which was estimated by placing an artificial disk having a uniform intensity of $2\sigma$ with 1$\sigma$ noise in the  PSF-subtracted image. 

Figure~\ref{fig:rotsub}, \ref{fig:psfsub}, and \ref{fig:med_locisub} show that there is a point-like emission (emission~B) very close to the location of HR~8799b reported by \cite{mar08}. The emission is faint, but we confirmed the detection of the emission~B based on the four different subtraction methods described in the previous section. The subtractions of the self-rotated images and self-median-filtered images also demonstrate that the detection is not an artifact by the mismatch of the PSFs of HR~8799 and the reference star (see the left panels of Figure~\ref{fig:rotsub} and \ref{fig:med_locisub}). In addition, the emission~B was found both before and after the dithering, indicating that the emission~B is not a cluster of hot pixels (Figure~\ref{fig:psfsub}, the bottom panels). 
The FWHM of the emission~B was measured to be $0\farcs07\pm0\farcs02$, which is in good agreement with the FWHM estimated for the masked PSF core of HR~8799. 
Therefore, we conclude that the emission~B is likely to be a real point source, and it corresponds to HR~8799b. 

The projected separation of HR~8799b from the parent star was measurable thanks to the transmissive coronagraphic mask. The centroid positions were measured, and their separation is obtained to be $(\Delta \alpha, \Delta \delta)=(1\farcs481\pm 0\farcs023, 0\farcs919 \pm 0\farcs017)$. The uncertainty includes the error of the plate scale and the 
uncertainty of the centroid measurements estimated based on 
the self-rotated image subtraction as described in the previous section. 
Since the centroid uncertainty derived in the DEC direction is unreasonably small (0.02~pixel), we further assumed the additional 0.5~pixel error. Those are the dominant error sources than the variation between the subtraction methods (PSF star subtraction, median-filtering, and the LOCI subtraction) and the frame selection (before or after the dithering). The possibility of a background object is clearly rejected as in \cite{mar08}: the proper motion of HR~8799 suggests that the relative separation of a background star to HR~8799 should be changed by $0\farcs71$ in 6 years, while it changed only $0\farcs02$ from July 2002 to 2008. 
The multi-epoch data are summarized in Table~\ref{tab:separation}. Our 2002 result also shows that HR~8799b was orbiting in counter-clockwise direction  (Figure~\ref{fig:orbit}) at a rate of $18^{+13}_{-9}$~mas~yr$^{-1}$ from 2002 to 2004. 
The projected separation $1\farcs743 \pm 0\farcs029$ corresponds to 68.7~AU assuming the distance of 39.4~pc. This is consistent with those in 2004--2008 within the uncertainty, indicating that the nearly circular orbit inferred by \cite{mar08} is plausible. 
 

The aperture photometry was made using the IRAF/APPHOT package with an aperture of 3~pixels in radius and the sky annulus of 5--10~pixels. The median-filtered images were not used for the photometry. The comparison with the photometry for HR~8799 may give a reasonable magnitude, but the transmission of the occulting mask has not been accurately determined, ranging from 1--2\% based on the measurements during the commissioning observations, to 11\% obtained with the laboratory experiment. Nevertheless, assuming the observed transmission of $\sim$1\%, the 3 pixel aperture photometry for the masked HR~8799 suggests $H = 18$~mag for HR~8799b. Alternatively, the aperture correction from 3 to 6 pixels (approximately 4~HWHM) ($\Delta H$ = 0.26~mag) was applied, which was derived from the artificial PSF modeled with a moffat function that fit the transmitted core of HR~8799. The flux uncertainty was estimated by considering the standard deviation of the photometric results for several data subsets as well as the uncertainty associated with the aperture correction, and the photometric errors for the flux calibrator and HR~8799b itself. In addition, the difference of the photometric results using the reference star PSF and the self-rotated images was considered. The inferred magnitude is $H=18.07\pm0.25$~mag, which is in agreement with that measured by \cite{mar08}. 
The agreement with \cite{mar08} suggests that the planet is not significantly variable in the interval of 6 years at its age (30--160~Myr; \citealt{mar08}). In addition, the consistency further supports their derivation of the planetary mass.

There cannot be seen other significant detections of point sources in the PSF-subtracted images due to the severe contamination by the subtraction residuals. We have to say that it is completely impossible to detect HR~8799d which is the closest one to the star ($r=0\farcs6$). There can be seen an emission at the expected location of HR~8799c ($r\sim1\arcsec$), but it is strongly affected by the speckle noise that cannot be removed at such radii. HR~8799c is therefore not detected. 

\section{Summary}
\begin{figure}
\plotone{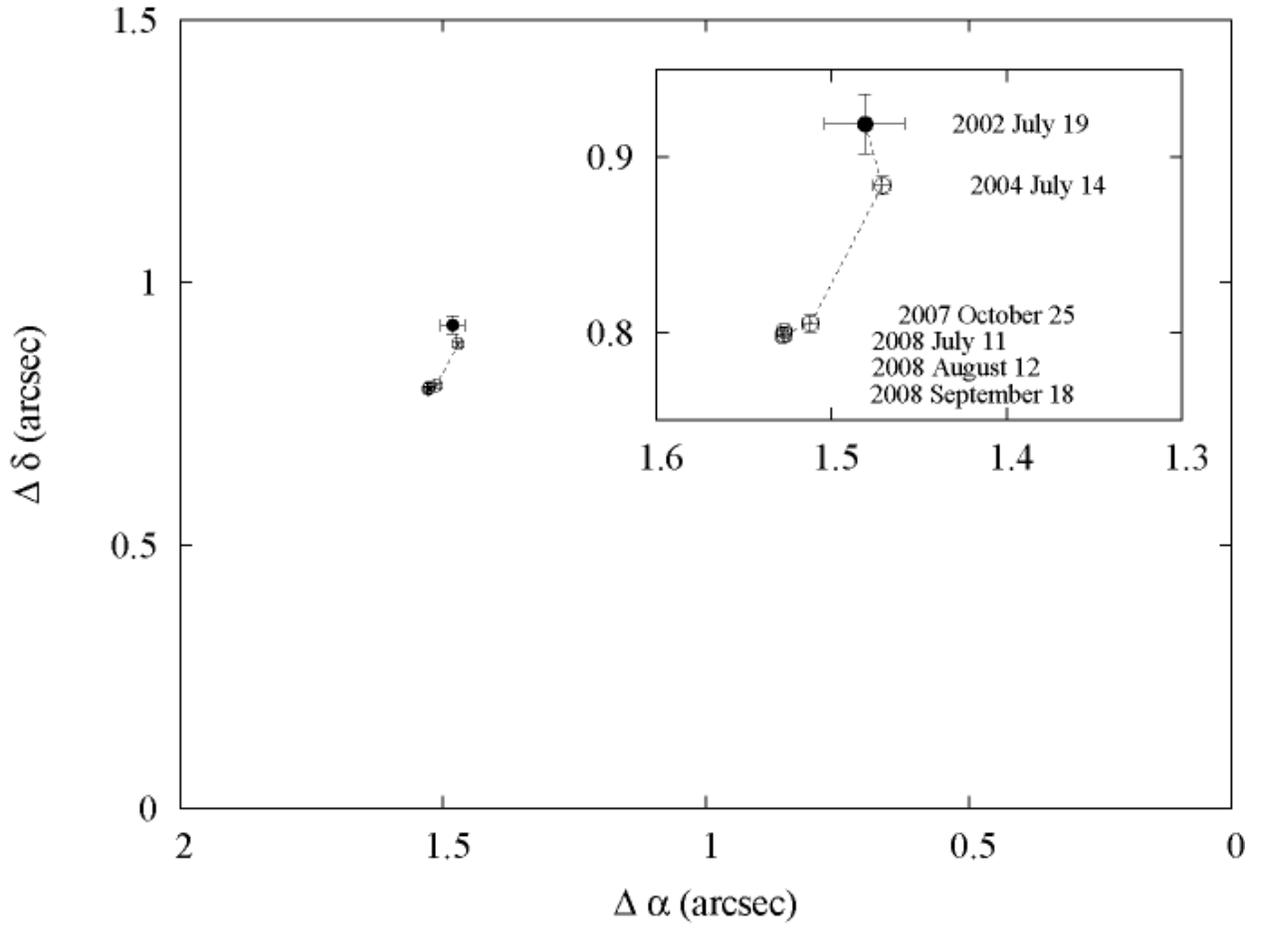}
\caption{Orbital motion of HR~8799b from 2002 to 2008. The positions in 2004--2008 were measured by \cite{mar08}. \label{fig:orbit}} 
\end{figure}

The imaging observations with the 36-elements AO revealed a planetary companion HR~8799b which has been discovered by \cite{mar08}. Our Subaru data obtained in July 2002 gave the longest time coverage for the positional measurement when combined with the Keck results. It is confirmed that HR~8799b is bound to the host star and orbiting counter-clockwise. Note that these data points do not produce a smooth track (Figure~\ref{fig:orbit}) and additional observations are required to derive the orbital solution. The photometric result is also consistent with the observations at later epochs. 

\acknowledgments
We appreciate the support from the Subaru Telescope staff, especially from K. Murakawa, H. Suto, and S. Harasawa. We are grateful to the referee for  valuable remarks and suggestions. YI is supported by the Global Centers of Excellence (GCOE) Program : "Foundation of International Center for Planetary Science" from MEXT. MT is supported by Grants-in-Aid for Scientific Research on Priority Areas, ``Development of Extra-Solar Planetary Science'' from the Ministry of Education, Culture, Sports, Science, and Technology (16077101, 16077204).


{\it Facilities:} \facility{Subaru (CIAO, AO36)}




\clearpage



\clearpage

\clearpage

\clearpage

\clearpage



\begin{deluxetable}{llll} 
\tablecolumns{5} 
\tablewidth{0pc} 
\tablecaption{Projected Separation for HR~8977b from the Parent Star\label{tab:separation}} 
\tablehead{\colhead{date} & \colhead{$\Delta \alpha$ ($\arcsec$)} & \colhead{$\Delta \delta$ ($\arcsec$) } & \colhead{reference}}
\startdata
2002 Jul. 19  & 1.481 $\pm$ 0.023 &  0.919 $\pm$ 0.017 & this work \\
2004 Jul. 14  & 1.471 $\pm$ 0.005 & 0.884 $\pm$ 0.005 & \cite{mar08} \\
2007 Oct. 25  & 1.512 $\pm$ 0.005 & 0.805 $\pm$ 0.005 & \cite{mar08} \\
2008 Jul. 11  & 1.527 $\pm$ 0.004 & 0.799 $\pm$ 0.004 &  \cite{mar08} \\
2008 Aug. 12  & 1.527 $\pm$ 0.002 & 0.801 $\pm$ 0.002 &  \cite{mar08} \\
2008 Sept. 18 & 1.528 $\pm$ 0.003 & 0.798 $\pm$ 0.003 & \cite{mar08} \\
\enddata
\tablecomments{The uncertainties of the plate scale and orientation for Keck/NIRC2 are well below the errors quoted in the Table \citep{mar08}. }
\end{deluxetable}


\end{document}